\documentclass[fleqn,twoside]{article}
\usepackage{espcrc2}
\usepackage{amsmath}
\usepackage{epsfig}

\usepackage{graphicx}
\usepackage[figuresright]{rotating}

\input babarsym

\newcommand{\etal}{{\em et al.}}

\title{Measurement of $\boldmath{\sin{2\beta}}$ in tree-dominated 
       $\Bz$-decays and ambiguity removal}

\author{H. Lacker\address{Institute of Nuclear and Particle Physics, 
        Technical University Dresden, D-01062 Dresden, Germany} \\
On behalf of the BaBar and Belle Collaborations
}
       
\begin{document}

\begin{abstract}
The most recent results from the $B$-factories on the time-dependent 
$CP$ asymmetries measured in $B^{0}$-decays mediated by $b \to c \bar{c} s$ 
quark-transitions are reviewed. The Standard Model interpretation of the results in terms 
of the parameter $\sin{2\beta}$ leads to a four-fold ambiguity on the unitarity 
triangle $\beta$ which can be reduced to a two-fold ambiguity by measuring the 
sign of the parameter $\cos{2\beta}$. The results on $\cos{2\beta}$ obtained 
so far are reviewed.
\vspace{1pc}
\end{abstract}

\maketitle

\section{INTRODUCTION}

Within the Standard Model, quark flavor-mixing is described by the $3 \times 3$ 
unitary Cabibbo-Kobayashi-Maskawa (CKM) matrix~\cite{CKM}. The size of \CP\
violation carried by the CKM matrix is quantified by the imaginary part 
of the apex, $\rhobar+i \etabar=-\frac{V_{ud}V_{ub}^{*}}{V_{cd}V_{cb}^{*}}$,
of the unitarity triangle. 
Constraints on the parameter \stwob\ 
($\beta ={\rm arctan}{(\frac{\etabar}{1-\rhobar})}$) can be obtained from 
the measurement of time-dependent decay rates of neutral $B$-mesons into $CP$ 
eigenstates $f_{CP}$ mediated by $b \to c \bar{c} s$ quark transitions.
The time-dependent rate asymmetry between $\Bzb$ and $\Bz$ decays is given by 
$A_{CP}(t) = \frac{2 \cdot \Im{\lambda}}{1+{|\lambda|}^2} \sin{(\Delta m_{d} \cdot t)} 
              - \frac{1-{|\lambda|}^2}{1+{|\lambda|}^2} \cos{(\Delta m_{d} \cdot t)}$,
where $\Delta m_{d}$ is the mass difference between the heavy ($B_{H} = p \Bz + q \Bzb$) 
and the light ($B_{L} = p \Bz - q \Bzb$) neutral mass eigenstate, and 
$S= \frac{2 \cdot \Im{\lambda}}{1+{|\lambda|}^2}$ and 
$C=-A=\frac{1-{|\lambda|}^2}{1+{|\lambda|}^2}$ where the parameter $C$ is
used by the \babar\  collaboration and $A$ by the Belle collaboration. 
The parameter $\lambda$,
given by 
$\eta_{CP} \frac{q}{p} \frac{\bar{A}_{{\bar{f}}_{CP}}}{A_{{f}_{CP}}}$,
 $\eta_{CP}$ 
being the $CP$ eigenvalue of the final state, is a phase-convention-independent 
quantity which contains possible sources of $CP$ violation in mixing 
($|\frac{q}{p}| \neq 1$), direct $CP$ violation
($|\frac{\bar{A}_{{f}_{CP}}}{A_{{f}_{CP}}}| \neq 1$), and $CP$ violation in the 
interference between decay with and without mixing ($\Im{\lambda} \neq 0$).
In the $\Bz-\Bzb$ system, to a very good approximation 
$\frac{q}{p} = \frac{V_{td}V_{tb}^{*}}{V_{td}^{*}V_{tb}}$ and $CP$-violation in 
mixing is expected to be small, which is confirmed by the experimental constraints: 
$|\frac{q}{p}| - 1 = 0.0015 \pm 0.0039$~\cite{HFAG}. 
For the decay $\Bz/\Bzb{\to} (c \bar{c}) K^{0}/{\bar{K}}^{0}$, 
the dominant amplitude is 
a tree-mediated $b \to c \bar{c} s$ quark-transition. There are also contributions 
from penguin amplitudes. In the case of the $c$- and $t$-penguin, the CKM phases 
are equal or almost equal, respectively, to the tree-decay phase 
($\propto V_{cs}V_{cb}^{*}$). Only in the case of the $u$-quark penguin the CKM 
phase ($\propto V_{us}V_{ub}^{*}$) differs significantly from the tree amplitude. 
However, it is doubly-Cabibbo suppressed with respect to the tree amplitude. 
An additional effect comes from $CP$-violation in the $K^{0}-{\bar{K}}^{0}$
system, however which changes the relative phase by only a small amount.
Hence, for the $CP$ asymmetry, one expects to a very good approximation
$A_{CP}(t)= \eta_{CP} \sin{2\beta} \sin{(\Delta m_{d} \cdot t)}$.

\section{MEASUREMENTS OF \boldmath$\sin{2\beta}$}

The most precise determinations of $\sin{2\beta}$ are coming currently from the 
$B$-factory experiments \babar\  and Belle thanks to the excellent performance
of their accelerators PEP II and KEKB. The new peak luminosity records achieved
in 2006 are $1.207 \times 10^{34}~{\rm cm^{-2} s^{-1}}$ for PEP II and  
$1.6517 \times 10^{34}~{\rm cm^{-2} s^{-1}}$ for KEKB.
By summer 2006, PEP II/\babar\ had collected $390~{\rm fb^{-1}}$ of 
data while KEKB/Belle had collected a data sample of $630~{\rm fb^{-1}}$. This results
in a sum of $10^{9}$ 
$B \bar{B}$ events written to tape. Both 
experiments have updated their $\sin{2\beta}$ analyses recently~\cite{Babarsin2b,Bellesin2b}.
 
At the $B$-factories, the $\Upsilon(4S)$ resonance is produced in $e^{+}e^{-}$
collisions with asymmetric electron and positron beam energies resulting in a 
boost factor of $\beta \gamma = 0.56$ at PEP II and $\beta \gamma = 0.425$ at KEKB, 
respectively.  
The $\Upsilon(4S)$ decays in about $50\%$ of the cases to neutral $B$-meson pairs, 
produced in a coherent quantum state. The time-dependent $CP$ asymmetry is measured 
by determining the decay-time difference $\Delta t$ between the decay of one $B$
to a $CP$ eigenstate ($B_{CP}$) and the decay of the other $B$ meson ($B_{tag}$) 
to flavour-specific final states. These final states are not reconstructed exclusively 
and the flavor of $B_{tag}$ is determined (``tagged'') only on a statistical basis 
by determining the sign of the charge mainly from high-energy leptons, and kaons and  
low-energy pions mutually stemming from $D^{*}$ decays. 
The figure of merit for the tagging performance is given by the quality factor 
$Q=\sum_{i} \epsilon_{i} (1-2 \times w_{i})^{2}$, where $\epsilon_{i}$ is the
tagging efficiency and $w_{i}$ the mistag fraction in tagging category $i$. $Q$ 
enlarges the statistical uncertainty of the decay-rate asymmetry measurements and has been 
measured by \babar\  (Belle) to be $Q=(30.4 \pm 0.3)\%$ ($Q=(29.0 \pm 1.0)\%$).
The decay-time difference $\Delta t$ is estimated from the distance between the two 
$B$-decay vertices $\Delta z$ in the beam-direction $z$. Due  to the $\Upsilon(4S)$ 
boost, $\Delta z$ can be measured with sufficient precision by means of silicon vertex 
detectors. The average $\Delta z$ is of order $200~{\mu \rm m}$ at Belle and about 
$250~{\mu \rm m}$ at \babar. The $\Delta z$ resolution is dominated by the $B_{tag}$ 
vertex resolution and is of order $190~{\mu \rm m}$.


The most recent \babar\  analysis~\cite{Babarsin2b} was performed on a sample of 
$348\times 10^{6}$ $B \bar{B}$ events. The $CP$-odd final states 
($\eta_{CP}\!=\!-1$) used are $\Bz/\Bzb{\to} (c \bar{c}) K^{0}_{S}$ where 
$c \bar{c}=J/\psi \to e^{+}e^{-},\mu^{+}\mu^{-}, 
   \psi(2S)\to e^{+}e^{-},\mu^{+}\mu^{-},J/\psi\pi^{+}\pi^{-}$, 
    $\chi_{c1} \to J/\psi \gamma$, and $\eta_{c} \to K^{0}_{S} K^{+} \pi^{-}$. 
The $CP$-even final state $\Bz/\Bzb{\to} J/\psi K^{0}_{L}$ is also reconstructed.
Compared to the high-purity sample $\Bz/\Bzb{\to} (c \bar{c}) K^{0}_{S}$, this 
final state suffers from a significantly higher background level. 

\begin{figure}[htb]
\begin{center}
\psfig{file=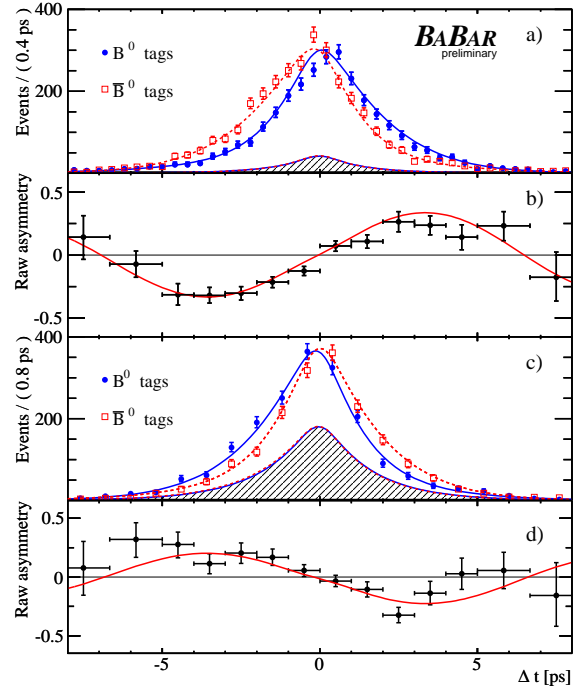,width=3.0in}
\end{center}
\caption{Measured decay-time distributions and corresponding decay-rate
asymmetries for the \babar~ $\sin2\beta$ analysis. 
a) and b) \CP\-odd finals states, c) and d) \CP\-even final state.
The shaded regions represent the estimated background contribution.
The solid (dashed) curves represent the fit projections in $\Delta t$
for $\B$ ($\Bb$) tags.}
\label{fig:babarsin2b}
\end{figure}

In addition, the vector-vector final state $\Bz/\Bzb \to J/\psi K^{*0}$ with 
$K^{*0} \to K^{0}_{S} \pi^{0}$ is used. A recent angular analysis performed by 
\babar\  finds an effective $CP$ eigenvalue of $\eta_{CP}=0.504\pm0.033$ for this final 
state~\cite{Babarjpsikstaramplitudes}. 
For the $CP$-odd final states the result
is $\sin{2\beta}=+0.713 \pm 0.038_{stat}$, for the final state 
$\Bz/\Bzb{\to} J/\psi K^{0}_{L}$ it is $\sin{2\beta}=+0.716 \pm 0.080_{stat}$
(see Fig.~\ref{fig:babarsin2b}). 
The average of all modes including $\Bz/\Bzb \to J/\psi K^{*0}$ results in  
$\sin{2\beta}=+0.710 \pm 0.034_{stat} \pm 0.019_{sys}$.
The main systematic uncertainties are due to the $CP$ content of the background,
the $\Delta t$ resolution functions, the background in $\Bz/\Bzb{\to} J/\psi K^{0}_{L}$,
the differences in the mistag fractions between $\Bz$ and $\Bzb$, and the knowledge
of the event-by-event beam spot position.
For the final state $\Bz/\Bzb {\to} (c \bar{c}) K^{0}_{S}$, which has the highest 
purity 
($92\%$), \babar\  fits $\sin{2\beta}$ and $|\lambda|$ simultaneously and finds 
$C=+0.070 \pm 0.028_{stat} \pm 0.018_{sys}$, translating into a $2~\sigma$ deviation 
from zero. The dominant systematic uncertainty in this case  ($0.014$) is due to the 
interference between the suppressed $\bar{b} \to \bar{u} c \bar{d}$ amplitude and 
the $\bar{b} \to c \bar{u} d$ amplitude for certain hadronic $B_{tag}$ decays.

The Belle analysis~\cite{Bellesin2b} uses a sample of $532$ million $B \bar{B}$ 
events where only the $CP$-odd final state $\Bz/\Bzb{\to} J/\psi K^{0}_{S}$ and the 
$CP$-even final state $\Bz/\Bzb{\to} J/\psi K^{0}_{L}$ are considered. Both results 
for the $S$ and $C$ coefficients are consistent: 
$\sin{2\beta}(J/\psi K^{0}_{S})=+0.643 \pm 0.038_{stat}$
and $\sin{2\beta}(J/\psi K^{0}_{L})=+0.641 \pm 0.057_{stat}$, respectively,
$C(J/\psi K^{0}_{S})=-0.018 \pm 0.021_{stat}$ and 
$C(J/\psi K^{0}_{L})=-0.045 \pm 0.033_{stat}$.
The averages between both results are:
$\sin{2\beta}=+0.642 \pm 0.031_{stat} \pm 0.017_{sys}$ and 
$C=-0.018 \pm 0.021_{stat} \pm 0.014_{sys}$.
The dominant systematic uncertainties on $\sin{2\beta}$ come from the vertex
reconstruction, the resolution function, the background fraction, flavor tagging,
and the effect of the tag-side interference.
The systematic uncertainty on $C$, as in the case of \babar, is
dominated by the tag-side interference ($0.009$).

The world average computed by the Heavy Flavor Averaging Group (HFAG)~\cite{HFAG}
is $\sin{2\beta}=+0.675 \pm 0.023_{stat} \pm 0.012_{sys}$, where most of the
systematic uncertainties have been treated as uncorrelated. Correlated error
sources like the $B$-lifetime, the mixing frequency $\Delta m_{d}$ or the effect
of the tag-side interference play only a minor role at this stage. This suggests 
that on the time scale of 2008, when an integrated luminosity of order 
$2~{\rm ab^{-1}}$ is expected from \babar\  and Belle together, the total uncertainty 
on $\sin{2\beta}$ will fall below $0.020$. 
Theoretical estimates for the difference between 
the measured $S$ coefficient and the true $\sin{2\beta}$ value are of order $0.01$ 
or below~\cite{sin2btheory}. 
The average value of \babar\  and Belle for the cosine 
coefficient is $C=0.012 \pm 0.017_{stat} \pm 0.014_{sys}$ and hence consistent with 
zero as expected in the Standard Model at this level of precision. The systematic 
uncertainty is dominated by the tag-side interference effect. Under the assumption 
that both measurements have estimated their uncertainties correctly and that the 
uncertainties are Gaussian-like, the p-value (denoted $CL$ by HFAG~\cite{HFAG}) to find a deviation 
between the two experiments as large as or larger than the one observed is 0.02.

\section{MEASUREMENTS OF \boldmath$\cos{2\beta}$}

Since only the parameter $\sin{2\beta}$ is measured, there are four different solutions 
for $\beta$. Two solutions correspond to $\cos{2\beta}>0$, $\beta=21.2^{\circ}$ or 
$\beta=21.2^{\circ}+180^{\circ}$, while the ones corresponding to $\cos{2\beta}<0$ 
are $\beta=68.8^{\circ}$ or $\beta=68.8^{\circ}+180^{\circ}$. The constraints on the 
unitarity triangle excluding $\sin{2\beta}$ as an input result is  
$\beta=(27.7^{+0.8}_{-3.9})^{\circ}$~\cite{PapII}, suggesting that $\beta=21.2^{\circ}$ 
is the correct solution. However, this argument relies on the validity of the 
Standard Model. If one allows for New Physics contributions in $\Bz-\Bzb$ mixing a 
new phase, $2\theta_{d}$, is introduced. As a result, the measured parameter is 
$\sin{(2\beta+2\theta_d)}$ and the new degree of freedom invalidates the argument given 
above. Only the measurement of a positive sign of $\cos{(2\beta+2\theta_d)}$ allows 
a restriction of the possible solutions to $\beta+\theta_{d}=21.2^{\circ}$ or 
$\beta+\theta_{d}=21.2^{\circ}+180^{\circ}$. Even then the allowed space for $\beta$ 
and $\theta_{d}$ is quite large and can only be reduced by adding other experimental 
inputs (see e.g. Ref.~\cite{PapII}).

\subsection{\boldmath$\Bz/\Bzb {\to}J/\psi K^{*0}$}

The first constraints on $\cos{2\beta}$ have been obtained from a time-dependent 
angular analysis of the final state $\Bz/\Bzb {\to}J/\psi K^{*0}$.
Since this is a vector-vector mode there are three amplitudes contributing to
this final state: two $CP$-even amplitudes, $|A_{0}|e^{i \delta_{0}}$ and 
$|A_{||}|e^{i \delta_{||}}$, and a $CP$-odd amplitude $|A_{\perp}|e^{i \delta_{\perp}}$,
where the $\delta_{i}$ represent strong phases. The sizes and relative strong 
phases of these amplitudes can be measured from the angular distributions 
(usually described in the transversity basis) of the decay products of the 
$J/\psi$ and $K^{*0}$.
The measured time-dependent decay-rate asymmetry in the three-dimensional phase 
space of the transversity angles is sensitive to $\cos{2\beta}$ due to the 
interference between the $CP$-even amplitudes $A_{0}$ and $A_{||}$, and the 
$CP$-odd amplitude $A_{\perp}$.
However, there is an ambiguity in the solution of the strong phase differences
$(\delta_{||}-\delta_{0},\delta_{\perp}-\delta_{0}) \to (\delta_{0}-\delta_{||},\pi+\delta_{0}-\delta_{\perp})$ resulting in a sign ambiguity 
$\cos{2\beta} \to -\cos{2\beta}$ which seems to spoil the $\cos{2\beta}$
extraction.

\babar\ has performed such a time-dependent angular analysis on a sample 
of $88\times 10^{6}$ $B \bar{B}$ pairs~\cite{BabarJPsiKstar} and finds
$\sin{2\beta}=+0.10 \pm 0.57_{stat} \pm 0.14_{sys}$ and
$\cos{2\beta}=+3.32 \pm ^{+0.76}_{-0.96 stat} \pm 0.27_{sys}$. When fixing
$\sin{2\beta}=0.731$ as their best measured value, \babar\  
finds $\cos{2\beta}=+2.72 \pm ^{+0.50}_{-0.79 stat} \pm 0.27_{sys}$.
The sign ambiguity in $\cos{2\beta}$ has been resolved in \babar's analysis
by taking advantage of the interference in the $K\pi$ system between the 
$P$-wave coming from the $K^{*0}$ decay and the underlying $S$-wave. 

Belle has also performed a time-dependent angular analysis on a sample of
$275 \times 10^{6}$ $B \bar{B}$ pairs~\cite{BelleJPsiKstar} and finds
$\sin{2\beta}=+0.24 \pm 0.31_{stat} \pm 0.05_{sys}$ and
$\cos{2\beta}=+0.56 \pm 0.79_{stat} \pm 0.11_{sys}$. When fixing 
$\sin{2\beta}=0.726$, $\cos{2\beta}=+0.87 \pm 0.74_{stat} \pm 0.12_{sys}$
is found. The sign ambiguity in $\cos{2\beta}$ has not been resolved.
Instead, Belle uses the theoretical argument of $s$-quark 
helicity conservation~\cite{SuzukiJPsiKstar} to select the solution for the 
strong phase difference. This solution is consistent with \babar's 
experimental result on the strong phase difference and leads to the positive 
sign for $\cos{2\beta}$.

\subsection{\boldmath$\Bz/\Bzb \to D^{(*)0}/\bar{D}^{(*)0} h^{0}$}

It has been proposed recently~\cite{Bondar} that the sign of $\cos{2\beta}$ can 
also be extracted from a time-dependent Dalitz plot analysis of the decay
$\Bz/\Bzb \to D^{(*)0}/\bar{D}^{(*)0} h^{0}$ where $h^{0}$ denotes a light neutral
meson. These color-suppressed decay topologies are mediated by tree diagrams. 
If one neglects doubly-Cabibbo suppressed diagrams, the leading relevant quark 
transitions are $\bar{b} \to \bar{c} u \bar{d}$ for $\Bz \to \bar{D}^{(*)0} h^{0}$, 
and $b \to c \bar{u} d$ for $\Bzb \to D^{(*)0} h^{0}$, respectively.
Interference between these amplitudes is obtained by reconstructing the neutral 
$D$ mesons in the common final state 
$D^{(*)0}/\bar{D}^{(*)0} \to K^{0}_{S}\pi^{+}\pi^{-} (\pi)$.
We assume no $CP$ violation in the $D^{(*)0}/\bar{D}^{(*)0}$ system and denote 
$f_{+-}=f(m_{K^{0}_{S}\pi^{+}}^{2},m_{K^{0}_{S}\pi^{-}}^{2})$ for the 
$\bar{D}^{(*)0}$ and $f_{-+}=f(m_{K^{0}_{S}\pi^{-}}^{2},m_{K^{0}_{S}\pi^{+}}^{2})$ for 
the $D^{(*)0}$ decay amplitudes, respectively. 
Then, the time-dependent $B$ decay amplitudes are given by:
\begin{eqnarray*}
M_{\Bzb}(t) &=& f_{-+} \cos{(\Delta m_d \cdot t/2)} \\
            &-& i e^{-2 i \beta}\eta_{h^{0}}(-1)^{\ell}f_{+-}\sin{(\Delta m_d \cdot t/2)} \\
M_{\Bz}(t)  &=& f_{+-} \cos{(\Delta m_d \cdot t/2)} \\ 
            &-& i e^{+2 i \beta}\eta_{h^{0}}(-1)^{\ell}f_{-+}\sin{(\Delta m_d \cdot t/2)}, 
\end{eqnarray*}
where also direct $CP$ violation as well as $CP$ violation in mixing in the neutral 
$B$ system has been neglected.
Here, $\eta_{h^{0}}$ is the $CP$ eigenvalue of $h^{0}$ and $\ell$ is the 
relative orbital angular momentum in the $D^{(*)} h^{0}$ system. Due to the 
interference between $f_{+-}$ and $f_{-+}$ over the Dalitz plane, the time-dependent 
decay-rate asymmetry allows us to extract simultaneously $\sin{2\beta}$ and 
$\cos{2\beta}$ once the model for the Dalitz plot amplitudes is fixed.

\begin{figure}[htb]
\begin{center}
\psfig{file=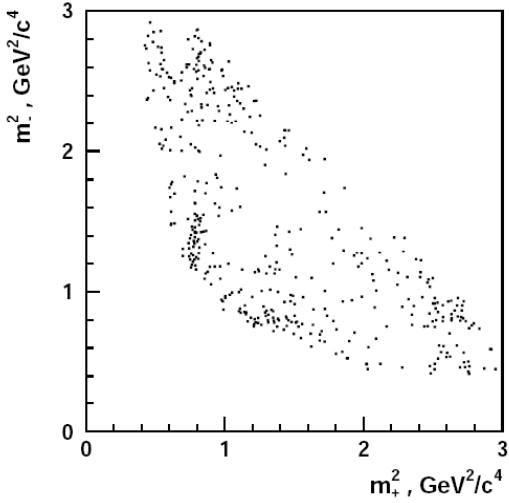,width=2.8in}
\end{center}
\caption{Dalitz plot distribution $m_{-}^2=m_{K^{0}_{S}\pi^{-}}^{2}$ versus
$m_{+}^2=m_{K^{0}_{S}\pi^{+}}^{2}$ as measured in the Belle 
$\Bz/\Bzb \to D^{0}/\bar{D}^{0} h^{0}$ analysis.}
\label{fig:dh0belledalitz}
\end{figure}

A time-dependent Dalitz plot analysis on a sample of $386$ million $B \bar{B}$ 
pairs has been first performed by Belle~\cite{Belled0h0} by reconstructing 
$\Bz/\Bzb \to D^{0}/\bar{D}^{0} h^{0}$ with $h^{0}=\pi^{0},\eta,\omega$
and $\Bz/\Bzb \to D^{*0}/\bar{D}^{*0} h^{0}$ with 
$D^{*0}/\bar{D}^{*0} \to D^{0}/\bar{D}^{0} \pi^{0}$ and $h^{0}=\pi^{0},\eta$.
The measured Dalitz plot is shown in Fig.~\ref{fig:dh0belledalitz}.
A simultaneous fit results in 
$\sin{2\beta}=+0.78 \pm 0.44_{stat} \pm 0.22_{sys+Dalitz}$ and
$\cos{2\beta}=+1.87^{+0.44}_{-0.53}(stat) ^{+0.22}_{-0.32}(sys+Dalitz)$.

In a similar \babar\ analysis on a sample of $311$ million $B \bar{B}$ 
pairs, the decays $\Bz/\Bzb \to D^{0}/\bar{D}^{0} h^{0}$ with 
$h^{0}=\pi^{0},\eta,\eta',\omega$ and 
$\Bz/\Bzb \to D^{*0}/\bar{D}^{*0} h^{0}$ with 
$D^{*0}/\bar{D}^{*0} \to D^{0}/\bar{D}^{0} \pi^{0}$ and
$h^{0}=\pi^{0},\eta$ have been 
reconstructed~\cite{Babard0h0}. \babar\  finds
$\sin{2\beta}=+0.45 \pm 0.36_{stat} \pm 0.05_{sys} \pm 0.07_{Dalitz}$ and
$\cos{2\beta}=+0.54 \pm 0.54_{stat} \pm 0.08_{sys} \pm 0.18_{Dalitz}$.
In this fit also $|\lambda|$ has been determined and found to be consistent 
with 1 ($0.98 \pm 0.09$). This is as expected due to the smallness of the 
neglected doubly-Cabibbo suppressed tree-amplitudes carrying a different 
CKM phase with respect to the dominant tree amplitudes.  In principle, the neglected amplitude could be taken into account in the 
analysis but this is likely to be impractical since the ratio of amplitudes
$\frac{|A(\Bzb \to \bar{D}^{0} h^{0})|}{|A(\Bzb \to D^{0} h^{0})|} \propto |\frac{V_{ub}V_{cd}^{*}}{V_{cb}V_{ud}^{*}}|$
is expected to be of order $0.02$. With higher statistics the precision on
$\cos{2\beta}$ might be dominated by the systematic uncertainty on the
Dalitz plot model which is already the largest systematic error.

\subsection{\boldmath$\Bz/\Bzb {\to}D^{*+} D^{*-} K^{0}_{S}$}

Another possible way to extract $\sin{2\beta}$ and $\cos{2\beta}$ 
simultaneously has been proposed in Ref.~\cite{Charles} and has then
been studied in more detail in Ref.~\cite{Browder}. The method uses the
final state $\Bz/\Bzb {\to}D^{*+} D^{*-} K^{0}_{S}$. The
time-dependent $CP$ asymmetry is given by:
\begin{eqnarray*}
A_{CP}(t;\sin{2\beta},\cos{2\beta}) = \eta_{y} \frac{J_{c}}{J_{0}} \cos{(\Delta m \cdot t)} \\
            - (\frac{2 J_{s1}}{J_{0}} \sin{2\beta} + \eta_{y} \frac{2 J_{s2}}{J_{0}} \cos{2\beta})\sin{(\Delta m \cdot t)}.
\end{eqnarray*}
The $J_{i}$ are integrals of functions of the amplitudes 
$A=A(\Bz \to D^{*+} D^{*-} K^{0}_{S})$ and $\bar{A}=A(\Bzb \to D^{*+} D^{*-} K^{0}_{S})$ 
over the half Dalitz space in the variables $s^{+}=m_{D^{*+}K^{0}_{S}}^{2}$ 
and $s^{-}=m_{D^{*-}K^{0}_{S}}^{2}$:
\begin{eqnarray*}
\begin{aligned}
J_{0(c)} & =  \int\limits_{s^{+}<s^{-}} (|A|^2 +(-) |\bar{A}|^2) ds,\\
J_{s1} & = \int\limits_{s^{+}<s^{-}} \Re{(\bar{A} A^{*})} ds,
J_{s2} = \int\limits_{s^{+}<s^{-}} \Im{(\bar{A} A^{*})} ds.
\end{aligned}
\end{eqnarray*}
The parameter $\cos{2\beta}$ can be measured if $J_{s2} \neq 0$. This could 
be realized if a broad intermediate resonance contributed. In this case, one
expects that $J_{c}$ becomes large.

\begin{figure}[htb]
\begin{center}
\psfig{file=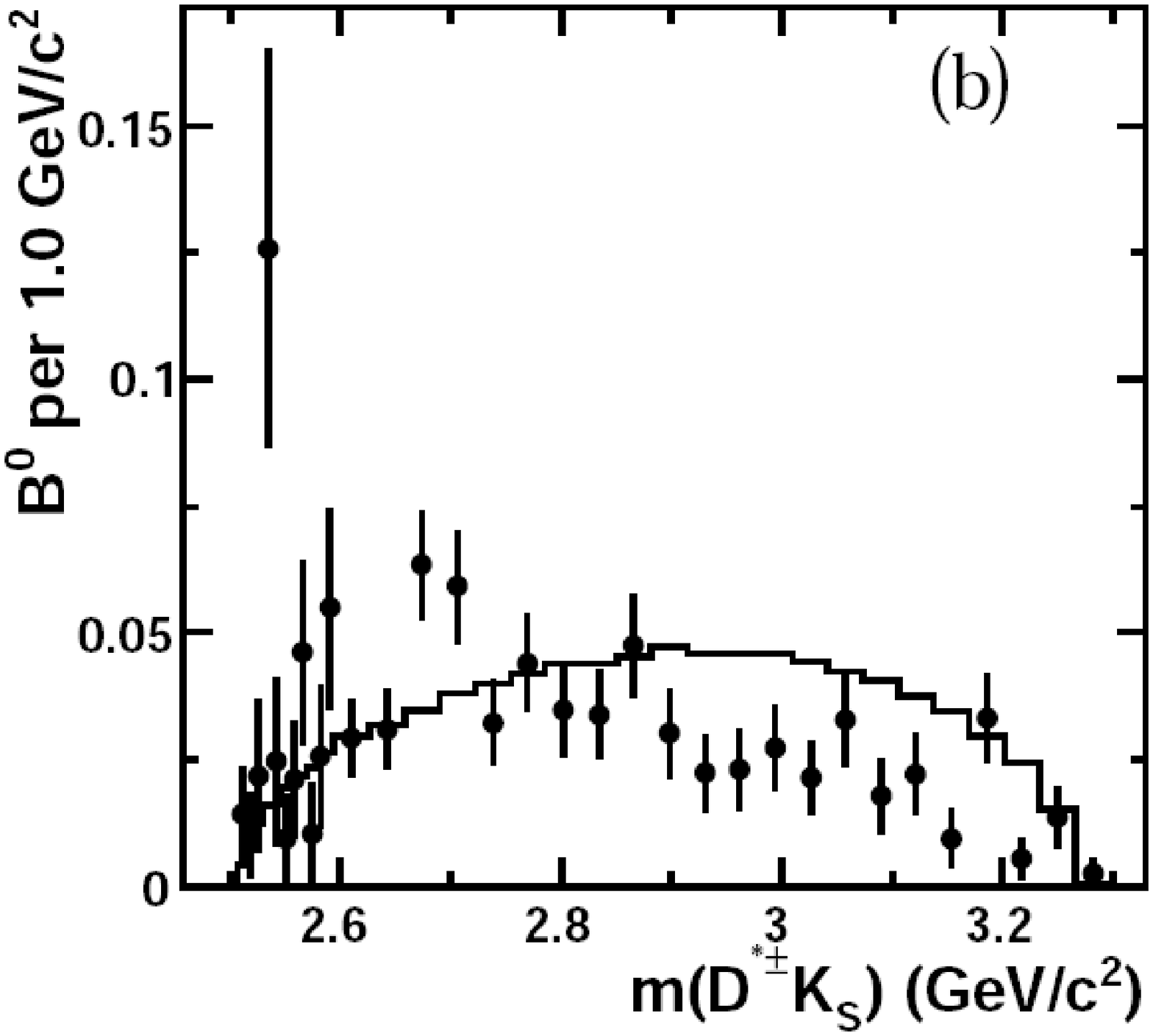,width=2.8in}
\end{center}
\caption{The $m_{D^{*\pm}K^{0}_{S}}$ distribution as measured in the \babar\ 
$\Bz/\Bzb {\to}D^{*+} D^{*-} K^{0}_{S}$ analysis.}
\label{fig:dsdsksmassbabar}
\end{figure}

The \babar\  collaboration has performed such an analysis~\cite{Babardsdsks}.
In the invariant $m_{D^{*\pm}K^{0}_{S}}$ spectrum there is a $4.6~\sigma$ evidence
for the $D_{s1}^{*}(2536)$ narrow width resonance. 
In addition, there is evidence for a broad structure below $2.9~{\rm GeV}/c^{2}$
(Fig.~\ref{fig:dsdsksmassbabar}). 
The results of the time-dependent analysis are 
$J_{c}/J_{0} = 0.76 \pm 0.18_{stat} \pm 0.07_{sys}$, being consistent 
with the presence of a broad resonance in the $D^{*\pm}K^{0}_{S}$ system, 
$2 J_{s1}/J_{0}\sin{2\beta} = +0.10 \pm 0.24_{stat} \pm 0.06_{sys}$, and 
$2 J_{s2}/J_{0}\cos{2\beta} = 0.38 \pm 0.24_{stat} \pm 0.05_{sys}$. 
In Ref.~\cite{Browder}, it has been predicted that the sign of $J_{s2}$ is
positive which would lead to a positive sign for $\cos{2\beta}$. 

An interesting question is the nature of the broad structure in the 
$m_{D^{*\pm}K^{0}_{S}}$ spectrum. Besides the $D_{s1}^{*}(2536)$ resonance a 
second $J^{P}=1^{+}$ state is expected. A possible candidate for this state
is the $D_{SJ}(2460)$ which however lies below the $D^{*\pm}K^{0}_{S}$ threshold.
In this case, the nature of this possible broad resonance is unclear.
\begin{table}[]
\caption{Confidence level to exclude $\cos{2\beta_{0}}<0$.}
\label{tab:cos2b}
\newcommand{\cc}[1]{\multicolumn{1}{c}{#1}}
\renewcommand{\arraystretch}{1.2} 
\begin{tabular}{@{}ccc}
\hline
Mode                                             & \babar\ & Belle     \\
\hline
$B {\to}J/\psi K^{*}$                     & $86\%$   & Not quantified  \\
$B \to D^{(*)0}/\bar{D}^{(*)0} h^{0}$     & $87\%$   & $98.3\%$  \\
$B {\to}D^{*+} D^{*-} K^{0}_{S}$          & $94\%$   & Not measured   \\
\hline
\end{tabular}
\end{table}

\subsection{Summary of \boldmath$\cos{2\beta}$ determinations}

The analyses reviewed here have not extracted a confidence level as a 
function of $\beta$.  
Instead, the $B {\to}J/\psi K^{*}$ \babar\  analysis~\cite{BabarJPsiKstar} 
has calculated a p-value for $\cos{2\beta_{0}}<0$, where $\beta_{0}$ corresponds 
to the value obtained from the the precise $\sin{2\beta}$ measurement. For $\cos{2\beta_{0}}<0$ the p-value 
is small ($0.6\%$). However, since the \babar-measured $\cos{2\beta}$ value lies 
outside the physical region, also the p-value for $\cos{2\beta_{0}}>0$ 
is small ($5.7\%$). For this reason, also the likelihood ratio 
$h_{-}/(h_{+}+h_{-})$ has been considered where $h_{\pm}$ is the probability 
density function value for the measured $\cos{2\beta}$ value if the true value is 
$\pm \cos{2\beta_{0}}$.
This likelihood ratio has then been interpreted in the framework of Bayesian 
statistics by assuming equal {\it a-priori} probabilities for the two hypotheses, 
$\cos{2\beta_{0}}<0$ and $\cos{2\beta_{0}}>0$, and denoted a confidence level.
The other analyses discussed in this review followed this procedure and the 
confidence levels obtained are summarized in Table~\ref{tab:cos2b}.

An average of all $\cos{2\beta}$ measurements is beyond the scope of this review. 
In some cases the results have central values outside the 
physical region and large asymmetric uncertainties which makes an average
cumbersome. The interpretation of the $\Bz/\Bzb {\to}D^{*+} D^{*-} K^{0}_{S}$ 
result relies on a theoretical assumption which requires further investigation, 
including the understanding of the Dalitz plot structure.
The systematic uncertainty due to the Dalitz plot model in
$\Bz/\Bzb \to D^{(*)0}/\bar{D}^{(*)0} h^{0}$ should be reduced since more 
statistics will come in the future. This channel provides a good  
opportunity to check whether $\sin{2\beta}$ is measured to be the same 
to a good approximation
in $b {\to}c \bar{c} s$ and $b {\to}c \bar{u} d$ transitions. 

Besides the above quantitative difficulties, the qualitative results suggest that the 
prefered solution is $\beta+\theta_{d}=21.2^{\circ}$ or 
$\beta+\theta_{d}=21.2^{\circ}+180^{\circ}$, in agreement with the Standard Model.
This finding removes half of the possible range in the New Physics 
parameter $2 \theta_{d}$.

\end{document}